\documentstyle{article}

\def\Journal#1#2#3#4{{#1} {\bf #2}, #3 (#4)}
\def\PRD{{\em Phys. Rev.} D}
\def\GRG{{\em Gen. Rel. Grav.}}
\def\CQG{{\em Class. Quant. Grav.}}
\def\PRS{{\em Proc. R. Soc. London} A}

\def\be{\begin{equation}}
\def\ee{\end{equation}}
\def\bea{\begin{eqnarray}}
\def\eea{\end{eqnarray}}

\begin{document}

\centerline{\bf Abrupt Changes in the Multipole 
Moments of a Gravitating Body}
\vskip 3truepc
\centerline{C. Barrab\`es and G. F. Bressange}
\centerline{Laboratoire de Math\'ematiques et Physique Th\'eorique,} 
\centerline{UPRES-A 6083 du CNRS,}
\centerline{Universit\'e de Tours, 37200 France;}
\vskip 1truepc
\centerline{and}
\vskip 1truepc
\centerline{P. A. Hogan}
\centerline{Mathematical Physics Department,}
\centerline{University College Dublin,}
\centerline{Belfield, Dublin 4, Ireland}

\vspace{2cm}

\abstract{ An  example is
described in    which  an asymptotically    flat static   vacuum  Weyl
space-time  experiences a sudden change  across a null hypersurface in
the multipole moments of   its  isolated axially symmetric  source.  A
light--like shell and an impulsive  gravitational wave are identified,
both having  the null hypersurface  as history.  The stress--energy in
the shell is dominated (at large distance from the source) by the jump
in the monopole moment (the mass) of  the source with  the jump in the
dipole moment mainly responsible for the stress being anisotropic. The
gravitational wave  owes its existence principally  to the jump in the
quadrupole moment of the source confirming what would be expected. 
This serves as a model of a cataclysmic astrophysical event such as 
a supernova. } 

\newpage
  
\section{Introduction}

Very few  examples are  known  describing gravitational  waves emitted
from an  isolated   source having its  wave-fronts  homeomorphic  to  a
2-dimensional sphere ("spherical radiation"). The principal example
is the Robinson-Trautman   family of solutions  \cite{RoTr}. These are
special, however,  because if the wave  fronts are sufficiently smooth
(no  conical singularities)   and  the  Riemann tensor   has   no line
singularity,  then these solutions  approach  Schwarzschild's solution
exponentially  in time  \cite{Luka}.  A   limiting  case is  Penrose's
spherical  impulsive   wave   propagating   through  flat   space-time
\cite{Pen}. More recent examples in the present context have also been
found by Alekseev   and Griffiths \cite{AlGr}.   When  these solutions
have the property that the Riemann tensor of the space-time containing
the   history of the  wave  involves a Dirac   delta function which is
singular  on   the null hypersurface  history  of  the  wave, then the
coefficient of the delta function is singular along a generator of the
null hypersurface.    This is known   as a  wire singularity  (or line
singularity).   The object of this  contribution is to  present a rare
example of a wire singularity-free impulsive gravitational wave 
in vacuum which is asymptotically  spherical  \cite{BBH}. 
To  construct  it, we  use the Barrab\`es-Israel formalism for 
light-like shells \cite{BI}. 

\section{Weyl static solutions in the Bondi form}
We consider the asymptotically flat Weyl static 
axially symmetric solutions \cite{EX} of Einstein's vacuum 
field equations in the Bondi form \cite{BBH}$^{\!,\,}$\cite{Bond}. 
	\be
		ds^2=-r^2\left\{f^{-1}d\theta ^2+f\,\sin ^2\theta\,d\phi ^2
		\right\}+2g\,du\,dr +2h\,du\,d\theta +c\,du^2\ ,
		\label{eq:le}
	\ee
with
	\bea
		f & = & 1-{Q\over r^3}\,\sin ^2\theta +
		O\left (r^{-4}\right )\ , \,\, \,\,g \ = \ 1+ 
O\left (r^{-4}\right )
\label{eq:cf}\\
		h & = & {2D\over r}\,\sin\theta +{3Q\over r^2}\,\sin\theta
		\,\cos\theta +O\left (r^{-3}\right )\ , \label{eq:ch}\\
		\hspace{-0.5cm}c & = & 1-{2m\over r}-{2D\over r^2}\,\cos\theta -
	{Q\over r^3}\left (3\cos ^2\theta -1\right )+O\left (r^{-4}\right )\ . 
\label{eq:cc}
	\eea
where the parameters $(m, D, Q)$ can be identified 
respectively as the mass, 
the dipole moment and the quadripole moment of the 
isolated source \cite{BBH}$^{\!,\,}$\cite{Bond}. 
In the form (\ref{eq:le}), the hypersurfaces $u=const.$ 
are exactly null \cite{BBH} (for all $r$ and not just for large $r$). 
Neglecting $O\left(r^{-4}\right)$-terms, the hypersurfaces $u=const.$ 
are generated by the integral lines of the futur-pointing 
null vector $\partial/\partial r$ and $r$ 
is an affine parameter along them. Moreover, 
these hypersurfaces are asymptotically future null cones \cite{BBH}.	

We subdivide the space--time $M$ with line--element (\ref{eq:le}) 
into two 
halves $M^-$ and $M^+$ having $u=0$ as common 
boundary. To the past of $u=0$, corresponding to $u<0$, 
the space--time $M^-$ is endowed with the metric (\ref{eq:le})
with parameters $\left\{m_-, D_-, Q_-,\dots\right\}$, metric 
coefficients 
$\{f, g, h, c\}$ given by  (\ref{eq:cf})-(\ref{eq:cc}) and 
coordinates $x^\mu _-=(\theta, \phi, r, u)$, 
while to the future of $u=0$, corresponding to $u>0$, 
the space--time $M^+$ is given by (\ref{eq:le}), metric 
coefficients 
$\{f_+, g_+, h_+, c_+\}$ given by  (\ref{eq:cf})-(\ref{eq:cc}) 
with 
parameters $\left\{m_+, D_+, Q_+,\dots\right\}$ and 
coordinates $x^\mu _+=(\theta _+, \phi_+, r_+, u)$. 
For convenience, we have taken $u_+=u$ and use $(\theta, \phi, r)$ 
as intrinsic coordinates on $u=0$. 
We now re-attach $M^-$ and $M^+$ on $u=0\,$ requiring only 
the continuity of the metric. From the perspective of the space--time 
$M^-\cup M^+$, we have an asymptotically flat Weyl solution $M^-$ 
undergoing an abrupt finite jump in its multipole moments across 
a null hypersurface $u=0$ resulting in the Weyl solution $M^+$. 
Applying now the Barrab\`es-Israel formalism \cite{BI}, 
we study the physical properties of the null hypersurface $u=0$. 
We find that it is the history of a light-like shell. 
In the system of coordinates of the $M^-$ side, 
the non-vanishing components of the stress-energy tensor 
of the shell are	 \cite{BBH}
	\bea
		S^{11} & = & O\left (r^{-7}\right )\ ,\hspace{1cm}S^{22}=
		O\left (r^{-7}\right )\ ,\nonumber\\
		16\pi\,S^{13} & = & -{3\left [D\right ]\over r^{4}}\,\sin \theta
		+O\left (r^{-5}\right )\ ,\nonumber\\
		16\pi\,S^{33} & = & -{4[m]\over r^2}-{12\left [D\right ]\over r^3}\,
		\cos\theta + {3\left [Q\right ]\over r^4}\,(5-11\,\cos ^2\theta )
		+O\left (r^{-5}\right )\ .
	\eea
where the square brackets denote the jump across $u=0$ of the 
quantity contained therein. The surface energy density of the 
shell measured by a radially moving observer is \cite{BBH}$^{\!,\,}$\cite{BI}
(up to a positive constant factor)
	\bea
	\sigma & := &-{1\over 4\pi r^2}\left\{[m]+{3\left [D\right ]\over r}\,
		\cos\theta -{3\left [Q\right ]\over 4r^2}\,(5-11\,\cos ^2\theta )
		+O\left (r^{-3}\right )\right\}\ . 
	\eea
It is natural to assume $[m] <0$ so the source suffers a loss of mass.
The results obtained in ref.\,\cite{BI} showed that the impulsive part 
of the Weyl tensor of the space-time $M^+\cup M^-$ splits into two parts 
which can be identified as a matter part and a wave part. 
This splitting has been carried out explicitly in ref.\,\cite{BBH}, 
demonstrating that in general, a light-like shell is accompanied by 
an impulsive gravitational wave which propagate independently along 
the null hypersurface $u=0$. Introducing a null tetrad asymptotically 
parallel transported along the integral curves of  
$\partial/ \partial r$, we find that the non-vanishing Newmann-Penrose 
components of the matter part of the delta function in the Weyl 
tensor are \cite{BBH}
	\be
	{}^M\Psi _3 \ = \ {3\sqrt {2}\left [D\right ]\over 4\,r^3}\,\sin\theta
	+O\left (r^{-4}\right )\,\,,\,\, {}^M\Psi _2=O\left (r^{-5}\right )
	\ee
It is predominantly  type III (with $n^\mu$ as 
degenerate principal null direction) in the Petrov classification because of
the anisotropy in the stress which in turn is  due to $[D]\neq 0$. 
The leading term of ${}^M\Psi _3$ has clearly no singularity 
for $0\leq\theta\leq\pi\nonumber$ and thus {\it no wire singularity}. The
only non-vanishing Newmann-Penrose component of the wave part is \cite{BBH}
	\be
	{}^W\Psi _4 =  -{3\left [Q\right  ]\over 4r^4}\,(3-7\,\cos ^2\theta)
		+O\left (r^{-5}\right )\ . 
	\ee
This impulsive gravitational wave is Petrov type N (with $n^\mu$ as 
four-fold degenerate principal null direction) and clearly owes its 
existence primarily to the jump in  the quadrupole moment  of 
the source across $u=0$  and also is manifestly {\it free of 
wire singularities}. Finally, analysing the induced metric on $u=0$, 
we see that, for a given r (which is asymptotically an affine 
parameter along the null geodesics of $u=0$), the corresponding 
2-surfaces are topologically spherical when neglecting 
$O\left(r^{-4}\right)$-terms. Hence the light--like shell and 
the impulsive gravitational 
wave can be considered asymptotically spherical in this sense.
We think that this example is a model of a cataclysmic astrophysical 
event such as a supernova which is likely to produce a burst of neutrinos 
travelling 
outward with the speed of light (modelled by the light--like 
shell) accompanied by a burst of outgoing gravitational waves 
(modelled by the impulsive wave). Another example of the coexistence of 
wire singularity-free spherical light-like shell and impulsive 
gravitational 
wave has been obtained by the same authors using a Kerr space-time 
undergoing a sudden change in the magnitude and in the direction 
of the angular momentum of the black-hole \cite{BBH2}.


\begin{thebibliography}{99}
\bibitem{RoTr}I. Robinson and A. Trautman, \Journal{\PRS}{265}{463}{1962}.
\bibitem{Luka}B. Lukacs, Z. Perj\`es, J. Porter and A. Sebestyen, 
\Journal{\GRG}{16}{691}{1984}.
\bibitem{Pen} R. Penrose in  {\it General Relativity: Papers  in
Honour  of  J. L. Synge}  ed. O'Raifeartaigh (Oxford: Clarandon Press)
p.101 (1972).
\bibitem{AlGr}G. A. Alekseev and J. B. Griffiths, 
\Journal{\CQG}{13}{L13}{1996}.
\bibitem{BBH}C. Barrab\`{e}s, G.F. Bressange and P.A. Hogan, 
\Journal{\PRD}{55}{3477}{1997}.
\bibitem{BBH2} C. Barrab\`{e}s, G.F. Bressange and P.A. Hogan, 
\Journal{\PRD}{55}{16}{1997}.
\bibitem{BI} C. Barrab{\`e}s and W. Israel, \Journal{\PRD}{43}{1129}{1991}.
\bibitem{EX} D. Kramer, H. Stephani, E. Herlt and M. MacCallum, {\it
Exact solutions     of Einstein's    field equations},  ed. E. Schmutzer
(Cambridge: Cambridge university press), p. 201 (1980).
\bibitem{Bond} H. Bondi, M. G. J. van der Burgh and A. W. K.  Metzner,
\Journal{\PRS}{269}{21}{1962}.
\end{thebibliography}
\end{document}